\documentclass{czjphyse}
\usepackage{epsf,rotate}

\newcommand{\inclfig}[2]{\mbox{\epsfysize=#1mm\epsffile{#2.eps}}}

\begin{document}

\title{Radiative corrections to deeply virtual Compton
                            scattering\footnote{Talk given by
			    D.\ M\"uller at Inter.\ Workshop ``PRAHA-Spin99'',
			    Prague, Sept.\ 6-11, 1999.}}
\authori{A.V. Belitsky$^a$, D. M\"uller$^b$, L. Niedermeier$^b$,
                            A. Sch\"afer$^b$}
\addressi{$^a$C.N.\ Yang Institute for Theoretical Physics,\\
                            State University of New York at Stony Brook\\
                            NY 11794-3800, Stony Brook, USA\\
	  $^b$Institut f\"ur Theoretische Physik, Universit\"at Regensburg\\
                            D-93040 Regensburg, Germany\\}
\authorii{}
%\authoriii{}
\headtitle{Radiative corrections to DVCS}
\headauthor{A.V. Belitsky et al.}  %use "et al." for more than 3 authors
%\specialhead{A.V. Belitsky et al.: Radiative corrections to DVCS}

%%%%%%%%%%%%%% FOR EDITORIAL USE ONLY!!! %%%%%%%%%%%%%%%
\evidence{A}
\daterec{XXX}    %;\\ final version }
\cislo{0}  \year{1999}
\setcounter{page}{1}
\pagesfromto{000--000}
%\makefirsttitle
%%%%%%%%%%%%%%%%%%%%%%%%%%%%%%%%%%%%%%%%%%%%%%%%%%%%%%%%

\maketitle

\begin{abstract}
We discuss possibilities of measurement of deeply virtual Compton scattering
amplitudes via different asymmetries in order to access the underlying
skewed parton distributions. Perturbative one-loop coefficient functions
and two-loop evolution kernels, calculated recently by a tentative use of
residual conformal symmetry of QCD, are used for a model dependent numerical
estimation of scattering amplitudes.
\end{abstract}

%%%%%%%%%%%%%%%%%%%%%%%%%%%%%%%%%%%%%%%%%%%%%%%%%%%%%%%%%%%%%%%%%%%%%%%%%%%%
\section{Introduction.}
%%%%%%%%%%%%%%%%%%%%%%%%%%%%%%%%%%%%%%%%%%%%%%%%%%%%%%%%%%%%%%%%%%%%%%%%%%%%

The asymptotic freedom of QCD allows to calculate short distance quantities
perturbatively form the first principles. Together with the use of
factorization theorems, the cross sections for (semi-)inclusive reactions
are predicted at a large scale as a convolution of perturbative parton
cross sections and long distance parton distributions. In the case of
exclusive reactions, such as two-photon reactions in the light-cone
dominated region or diffractive processes, the amplitudes factorizes in
hard-scattering amplitudes and hadron distribution amplitudes and/or
skewed parton distributions (SPD). Such distributions were already proposed
in the past \cite{MueRobGeyDitHor94} and they become more popular nowadays
\cite{Ji97Rad96ColFraStr96} since it has been realized their vital r\^ole
for the proper description of the mentioned reactions. Moreover, as
it was pointed out by J.\ Xi, their measurement can clarify the issue of
how the hadron total angular momentum is distributed among its
constiutents. One can expect that the study of the SPDs will be one of the
central issues of hadronic physics in nearest future.

A promising way to access the SPDs is to probe the nucleon with a highly
virtual photon via the Compton scattering process, which is called deeply
virtual Compton scattering (DVCS). Indeed, it is auspicious that a first
evidence of experimental feasibility to detect such a signal has been
revealed by the ZEUS collaboration \cite{Sau99}. On the theoretical side
the factorization of the collinear singularities has been proved at leading
twist-two level to all orders in coupling \cite{Rad97ColFre98JiOsb98}.
The local version of the OPE does not converge in the considered
kinematical region. Fortunately, this problem can be overcome by a
non-local light-cone expansion in terms of light-ray operators
\cite{AniZav78GeyRobBorHor85BalBra89,MueRobGeyDitHor94}. Moreover, from
the renormalization group equation of these operators one can derive in
a straightforward manner the evolution equations for SPDs.

Here we shortly review our approach to complete next-to-leading (NLO)
calculation, which is accomplished by a heavy use of the conformal
symmetry to map the known results from forward deep inelastic scattering
(DIS) to the non-forward kinematics of DVCS and simple one-loop addenda.
The problems within our modus operandi have been resolved by employing
conformal Ward identities and constraints arising from the algebra of
the collinear conformal group \cite{Mue94BelMue98aBelMue98c}.

%%%%%%%%%%%%%%%%%%%%%%%%%%%%%%%%%%%%%%%%%%%%%%%%%%%%%%%%%%%%%%%%%%%%%%%%%%%%
\section{The virtual Compton scattering process.}
%%%%%%%%%%%%%%%%%%%%%%%%%%%%%%%%%%%%%%%%%%%%%%%%%%%%%%%%%%%%%%%%%%%%%%%%%%%%

\begin{figure}[htb]
%\unitlength1mm
\centering
\inclfig{40}{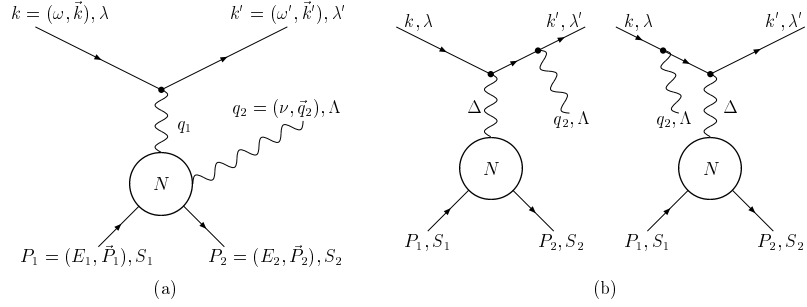}
%\begin{picture}(120,35)(0,0)
%\put(0,-8){\inclfig{12}{dvcsbh}}
%\end{picture}
\caption{The virtual Compton scattering process (a) and
the Bethe-Heitler process (b).}
\label{DVCS&BH}
\end{figure}

The virtual Compton scattering (VCS) process on a nucleon can be measured
in the reaction
\begin{equation}
\ell^\pm N \to \ell^\pm N \gamma.
\end{equation}
However, as shown in Fig.\ \ref{DVCS&BH}, there is contamination from the
so-called Bethe-Heitler (BH) process in which the final real photon is
radiated from the lepton beam.
%%%%%%%%%%%%%%%%%%%%%%%%%%%%%%%%%%%%%%%%%%%%%%%%%%%%%%%%%%%%%%%%%%%%%%%%%%%%
%                               Figure 1
%%%%%%%%%%%%%%%%%%%%%%%%%%%%%%%%%%%%%%%%%%%%%%%%%%%%%%%%%%%%%%%%%%%%%%%%%%%%
Thus, the scattering amplitude square $|T|^2$ in the cross section,
\begin{eqnarray}
\label{cross-section-def}
d \sigma = \frac{1}{4 k P_1} |T|^2
(2\pi)^4 \delta^4( k + P_1 - k^\prime - P_2 - q_2 )
\frac{d^3 \vec{k}^\prime}{2 \omega^\prime (2\pi)^3}
\frac{d^3 \vec{P}_2}{2 E_2 (2 \pi)^3}
\frac{d^3 \vec{q}_2}{2 \nu (2 \pi)^3},
\end{eqnarray}
contains beside the VCS and BH parts also the interference term:
\begin{eqnarray}
|T|^2 = \sum_{\lambda^\prime, S_2, \Lambda}
\left\{
|T_{VCS}|^2 + |T_{BH}|^2 + T_{VCS} T_{BH}^\ast + T_{VCS}^\ast T_{BH}
\right\}.
\end{eqnarray}
The BH-amplitude is purely real and is given as a contraction of the
leptonic tensor
\begin{equation}
L_{\mu\nu}
= \bar u (k^\prime, \lambda^\prime)
\left[
\gamma_\mu (\slash\!\!\! k - \slash\!\!\!\!\Delta)^{-1} \gamma_\nu
+ \gamma_\nu (\slash\!\!\! k^\prime + \slash\!\!\!\!\Delta)^{-1} \gamma_\mu
\right]
u(k, \lambda)
\end{equation}
with the hadronic current $J_\nu = \bar u (P_2)\left\{ F_1(\Delta^2)
\gamma_\nu + i F_2(\Delta^2) \sigma_{\nu\tau} \frac{\Delta^\tau}{2M}
\right\} u(P_1)$:
\begin{eqnarray}
\label{BH-amplitude}
T_{BH} = - \frac{e^3}{\Delta^2} \epsilon^\ast(\Lambda)_\mu L^{\mu\nu} J_\nu,
\quad \mbox{where}\quad \Delta = P_1 - P_2.
\end{eqnarray}
Since Dirac and Pauli form factors $F_1$ and $F_2$ are (partially) known
from experimental measurements and can be parametrized to a certain accuracy
by double dipole formulae, one can disentangle the content of the hadronic
tensor $T_{VCS}$:
\begin{eqnarray}
\label{VCS-amplitude}
T_{VCS} = \mp \frac{e^3}{q_1^2} \epsilon^\ast (\Lambda)_\mu T^{\mu\nu}
\bar u (k^\prime) \gamma_\nu u(k),
\mbox{\ where\ }
\left\{ {- \mbox{\ for\ } e^- \atop + \mbox{\ for\ } e^+ } \right. .
\end{eqnarray}
This tensor, defined by the time ordered product of two electromagnetic
currents
\begin{eqnarray}
\label{def-hadten}
T_{\mu\nu} (P, q_1, q_2)
= i \int dx e^{ix(q_1 + q_2)}
\langle P_2, S_2 |
T \left\{ j_\mu ( x/2 ) j_\nu( - x/2 ) \right\}
| P_1, S_1 \rangle,
\end{eqnarray}
contains for a spin-1/2 target twelve\footnote{$12 = \frac{1}{2} \times
3 \mbox{\ (virtual photon)} \times 2 \mbox{\ (final photon)} \times
2 \mbox{\ (initial nucleon)} \times 2 \mbox{\ (final nucleon)}$. The
reduction factor $1/2$ is a result of parity invariance.} independent
kinematical structures \cite{KroSchGui96}.

Measuring the cross section (\ref{cross-section-def}) in quite different
experiments: with negatively or positively charged lepton beam, with
(un)polarized beam or (un)polarized target, the rich structure of the
hadronic tensor can be explored in different kinematical regions. In the
following we are interested in DVCS, where $-q^2_1$ is large and $\Delta^2$
is small. Unfortunately, since the DVCS amplitude (\ref{VCS-amplitude})
decreases with increasing $q^2_1$ while the BH amplitude (\ref{BH-amplitude})
grows up with decreasing $\Delta^2$, in this kinematics it will be a
delicate task to extract the VCS amplitude for large virtualities $-q_1^2$
and especially low $\Delta^2$ from experimental data. On the other hand it
is promising to look for asymmetries which allow one to measure the
interference term directly and get in this way access to the real and
imaginary part of the DVCS amplitude \cite{FraFreStr97GuiVan98FreStr99}:
\begin{enumerate}
\item \underline{charge asymmetry}
\begin{eqnarray}
d \sigma(e^- \mbox{-beam}) - d \sigma(e^+ \mbox{-beam}) ,
\propto 4 {\rm Re} T_{\rm DVCS} T_{\rm BH}
\end{eqnarray}
\item \underline{single spin asymmetries} (either the lepton-beam or the
nucleon is polarized)
\begin{eqnarray}
d \sigma(\uparrow) - d \sigma(\downarrow)
\propto	4 {\rm Im} T_{\rm DVCS} T_{\rm BH},
\end{eqnarray}
\item \underline{azimuthal angle asymmetry}
\begin{eqnarray}
A = \frac{\int_{-\pi/2}^{\pi/2} d \phi_r d \sigma
- \int_{\pi/2}^{3\pi/2} d \phi_r d \sigma}{
\int_{0}^{2\pi} d \phi_r d \sigma},
\end{eqnarray}
\end{enumerate}
where $\phi_r = \phi_N - \phi_e$ with $ \phi_N (\phi_e)$ is the azimuthal
angle of the final nucleon (lepton).

%%%%%%%%%%%%%%%%%%%%%%%%%%%%%%%%%%%%%%%%%%%%%%%%%%%%%%%%%%%%%%%%%%%%%%%%%%%%
\section{Practical implication of conformal symmetry.}
%%%%%%%%%%%%%%%%%%%%%%%%%%%%%%%%%%%%%%%%%%%%%%%%%%%%%%%%%%%%%%%%%%%%%%%%%%%%

To describe different ``two-photon'' processes in the light-cone dominated
region in a unique way, we generalize the Bjorken limit and introduce the
scaling variables:
\begin{eqnarray}
\label{def-KinVar}
\xi \equiv \frac{Q^2}{P q},
\quad
\eta \equiv \frac{\Delta q}{P q},
\quad
Q^2 = -q^2 = - \frac{1}{4} ( q_1 + q_2 )^2,
\quad
P = P_1 + P_2,
\end{eqnarray}
where $\xi\approx x_{\rm Bj}/(2 - x_{\rm Bj}) $ is related to the
Bjorken variable $x_{\rm Bj} = - q^2_1/2(P_1 q_1)$ and $\eta$ is the
so-called skewedness parameter. Formally, this parameter interpolates
between different processes as given in table \ref{tab-KinLCd}.
\begin{table}[hbt]
\begin{center}
\begin{tabular}{|l|c|l|}
\hline
DIS              & $\Delta = 0$  & $\quad\eta  = 0$           \\
DVCS             & $q_2^2=0$     & $\quad \eta = 1/\omega$    \\
$\gamma^\ast N \to Nl^+ l^-$
                 & $q_2^2>0$     & $\quad\eta \approx \cos\phi_{\rm Br}
		                   = {{\bf p}_1{\bf Q}/|{\bf p}_1||{\bf Q}|}$,
                                   Breit--frame               \\
$\gamma^\ast\gamma^\ast \to H H$
                 &               & $\quad\eta \approx \cos\phi_{\rm cm}
				   = {{\bf p}_1{\bf Q}/|{\bf p}_1||{\bf Q}|}$,
                                   center-of-mass frame       \\
$\gamma^\ast \gamma^\ast \to M$
                 & $p_1=0$       & $\quad\eta=1$              \\
\hline
\end{tabular}
\end{center}
\vspace{-0.8cm}
\caption{\label{tab-KinLCd} The value of the scaling variable $\eta$
for different two-photon processes.}
\end{table}

Since in the case of DVCS the longitudinal part of the hadronic tensor
vanishes, there are only two kinematical structures left to leading twist-two
accuracy:
\begin{eqnarray}
\label{decom-T}
T_{\mu\nu}
= - \tilde g^T_{\mu\nu}
{\cal F}_1 ( \xi, \eta = \xi, Q^2, \Delta^2 )
+ i \tilde \epsilon_{\mu\nu q P} \frac{1}{Pq}
{\cal G}_1 ( \xi, \eta = \xi, Q^2, \Delta^2 )
+ \cdots.
\end{eqnarray}
Here the transverse part
of the metric tensor, denoted by ${g}^T_{\mu\nu}$, and the $\epsilon$-tensor
are contracted with the projection operators ${\cal P}_{\alpha\beta} =
g_{\alpha\beta} - q_{1\alpha} q_{2\beta} / (q_1 q_2)$:
$\tilde g^T_{\mu\nu} =
{\cal P}_{\mu}^{\ \alpha} {g}^T_{\alpha\beta} {\cal P}_{\ \nu}^{\beta}$,
$\tilde \epsilon_{\mu\nu q P} =
{\cal P}_{\mu}^{\ \alpha} \epsilon_{\alpha\beta q P}
{\cal P}_{\ \nu}^{\beta}$. The appearing amplitudes can be calculated by
means of the OPE. In contrast to the case of DIS, a larger number of
(local) operators and their Wilson coefficients contribute in the
non-forward case.  Fortunately, we can appreciate the conformal OPE (COPE)
\cite{FerGriGat71FerGriGat72a}, which tells us that all Wilson coefficients
are fixed up to an overall normalization, which can be taken from known
results in DIS. The conformal operators appearing in the COPE are:
\begin{eqnarray}
\left\{\!\!\!
\begin{array}{c}
{^Q\!{\cal O}^V} \\
{^Q\!{\cal O}^A}
\end{array}
\!\!\!\right\}_{jl}
&=&
\bar{\psi} (i \partial_+)^l\!
\left\{\!\!\!
\begin{array}{c}
\gamma_+ \\
\gamma_+ \gamma_5
\end{array}
\!\!\!\right\}
\!C^{\frac{3}{2}}_j\!
\left( \frac{\stackrel{\leftrightarrow}{D}_+}{\partial_+} \right)
\!\psi,
\\
\left\{\!\!\!
\begin{array}{c}
{^G\!{\cal O}^V} \\
{^G\!{\cal O}^A}
\end{array}
\!\!\!\right\}_{jl}
&=&
G_{+ \mu} (i \partial_+)^{l-1}\!
\left\{\!\!\!
\begin{array}{c}
g_{\mu\nu} \\
i\epsilon_{\mu\nu-+}
\end{array}
\!\!\!\right\}
\!C^{\frac{5}{2}}_{j - 1}\!
\left(
\frac{\stackrel{\leftrightarrow}{D}_+}{\partial_+}
\right)
\!G_{\nu +},
\nonumber
\end{eqnarray}
where $\partial \!= \stackrel{\rightarrow}{\partial}
\!\!+\!\! \stackrel{\leftarrow}{\partial}$
and  $\stackrel{\leftrightarrow}{D}
= \stackrel{\rightarrow}{D} - \stackrel{\leftarrow}{D}$. The $+$ and $-$
components are obtained by contraction with the two light-like vectors $n$
and $n^*$, such that $n^2 = n^{*2} = 0$ and $nn^* = 1$. They carry the spin
quantum number $l + 1$ and the conformal spin $j + 1$. In the conformal
subtraction (CS) scheme their mixing is only induced by the trace anomaly
and proportional to $\beta$ \cite{Mue97aBelMue97a}:
\begin{eqnarray}
\mu \frac{d}{d\mu} \mbox{\boldmath ${\cal O}$}_{jl}
= \mbox{\boldmath $\gamma$}_{jj}
\mbox{\boldmath ${\cal O}$}_{jl}
+ \frac{\beta}{g} \sum_{k = 0}^{j - 2}
\mbox{\boldmath $\Delta$}_{jk} \mbox{\boldmath ${\cal O}$}_{kl}
\end{eqnarray}
The anomalous dimension matrix $\mbox{\boldmath $\gamma$}_{jj}$ is, up to
a normalization factor in the mixed channels, the same as in DIS.

Unfortunately, for the case of DVCS kinematics the COPE in its local version
does not converge. Thus, it is necessary to resum the local operators into
so-called light-ray ones. This provide us a definition for the SPDs:
\begin{eqnarray}
\left\{ { {^Q\!q^V} \atop {^Q\!q^A} } \right\} ( t, \eta)
&=& \int \frac{d\kappa}{2\pi} e^{i \kappa t P_+}
\langle P_2 S_2 |
\bar \psi (- \kappa n)
\left\{ { \gamma_+ \atop \gamma_+ \gamma_5 } \right\}
\psi (\kappa n)
| P_1 S_1 \rangle ,
\\
\left\{ { {^G\!q^V} \atop {^G\!q^A} } \right\} ( t, \eta)
&=& \frac{4}{P_+} \int \frac{d\kappa}{2\pi} e^{i \kappa t P_+}
\langle P_2 S_2 |
G^a_{+ \mu} (-\kappa n)
\left\{ { g_{\mu\nu} \atop i \epsilon_{\mu\nu-+} } \right\}
G^a_{\nu+} (\kappa n)
| P_1 S_1 \rangle.
\nonumber
\end{eqnarray}
The momentum fractions of the incoming [outgoing] parton is
$\frac{t+\eta}{1+\eta} P_+$  [$\frac{t-\eta}{1+\eta} P_+$].
A form factor decomposition would give us the functions $H$, $E$ and
$\widetilde H$, $\widetilde E$ for the parity even and odd sectors,
respectively, as introduced by Ji \cite{Ji97Rad96ColFraStr96}.

Next we have to reconstruct the coefficient functions appearing in
\begin{eqnarray}
\label{def-T}
{\cal T}^a = \sum_{Q}
e^2_Q
\Bigg[
{^Q T^a} \otimes
{^Q\! q^a}
+ \frac{1}{N_f} \frac{1}{\eta}
{^G T^a} \otimes {^G\! q^a} \Bigg],
\end{eqnarray}
where ${\cal T}^V = {\cal F}_1$, ${\cal T}^A = {\cal G}_1$ and the sum runs
over quark species $Q = u, d, s$ with electrical charge $e_Q$.
The convolution is defined by $\tau_1 \otimes \tau_2 \equiv \tau_1 (x, y)
\tau_2 (y, z)$. This task can be achieved by some manipulation of the COPE
and, e.g.\ for the flavour non-singlet case, can be cast into the form
\begin{eqnarray}
\label{nonloc-T}
T (\xi, \eta, t, Q^2, \mu^2)
= \frac{1}{\xi}
F \left( \frac{\xi}{\eta}, \frac{r}{\eta} \right)
\otimes
\left( \frac{Q^2}{\mu^2} \right)^{
V \left( \frac{r}{\eta}, \frac{s}{\eta} \right) }
\otimes
C \left( \frac{s}{\eta}, \frac{t}{\eta}; \alpha_s \right).
\end{eqnarray}
The evolution kernel $V \left( \frac{r}{\eta}, \frac{s}{\eta} \right)$ and
the coefficient function $C \left( \frac{s}{\eta}, \frac{t}{\eta}; \alpha_s
\right)$ are diagonal w.r.t.\ the conformal waves --- the Gegenbauer
polynomials:
\begin{equation}
\label{coef-func}
\eta^{j} C_{j}^{3/2} \left( \frac{t}{\eta} \right)
\otimes
\left\{ { V \atop C }\right\}
\left( \frac{t}{\eta}, \frac{t'}{\eta}; \alpha_s \right)
=
\left\{ { - \frac{1}{2} \gamma_j (\alpha_s)
\atop
\phantom{- \frac{1}{2}} c_j (\alpha_s) }\right\}
\eta^j C_{j}^{3/2} \left( \frac{t'}{\eta} \right) ,
\end{equation}
and their eigenvalues $\gamma_j(\alpha_s)$ and $c_j(\alpha_s)$ coincide
with anomalous dimensions and Wilson coefficients, respectively, which
appear in DIS. A closed form of the function $F$ can be deduced order
by order in $\alpha_s$ from the sum
\begin{eqnarray}
\label{def-F}
F (x, y) &=& \sum_{j = 0}^\infty
\left( \frac{2 x}{ 1 + x } \right)^{j + 1}
\frac{B(j + 1, j + 2)}{(1 + x)^{\gamma_j/2}}  \\
&&\times {_2F_1} \left( \left.
{ 1 + j + {\gamma_j/2}, \ 2 + j + {\gamma_j/2} \atop 4 + 2j + {\gamma_j} }
\right| \frac{2 x}{1 + x} \right) C_{j}^{3/2} (y) .
\nonumber
\end{eqnarray}

Conformal covariance does not hold true in the minimal subtraction (MS)
scheme. The transformation, which rotates the CS result (\ref{nonloc-T}
-\ref{def-F}) to the minimal one, is induced by the conformal anomalies
appearing in the MS scheme. The rotation matrix can be found by study of the
conformal Ward identities for Green functions with operator insertions and
calculating the anomalous contributions via (modified) Feynman rules.
Indeed, the off-diagonal piece of the anomalous dimension matrix to NLO is
induced by the special conformal anomaly matrix $\mbox{\boldmath
$\gamma$}^{(0)c}$ and the first coefficient of the $\beta$-function
\cite{Mue94BelMue98aBelMue98c}:
\begin{eqnarray}
\label{adcodt}
\mbox{\boldmath $\gamma$}^{(1)}_{jk}
= \mbox{\boldmath $\gamma$}^{(1)}_{jj}
- {\left[
\mbox{\boldmath $\gamma$}^{(0)c} + \beta_0 \mbox{\boldmath $b$},
\mbox{\boldmath $\gamma$}^{(0)}	\right]_{jk}
\over 2 (j - k)(j + k + 3)}, \
\mbox{\boldmath $b$}_{jk}
= - \left[ 1 + (-1)^{j - k} \right]\theta_{jk} (2 k + 3)
\mbox{\boldmath $1$} .
\end{eqnarray}

The remaining task to transform the coefficient functions to the MS scheme
have been solved already in \cite{Mue97aBelMue97a}. In the case of DVCS,
they are nothing else as the analytical continuation of the hard-scattering
amplitudes for the meson production by two photon-collision. In the case of
the axial-vector channel they read:
\begin{eqnarray}
{^{Q}\!T}^{(0)}
\!\!\!&=&\!\!\!
\frac{1}{ 1 - t},
\qquad {^{G}\!T}^{(0)} = 0,
\\
{^{Q}\!T^{A(1)}}
\!\!\!&=&\!\!\!
\frac{C_F}{2 (1 - t)}
\Bigg[ \left( 2 \ln \frac{1 - t}{2} + 3 \right)
\left(
\frac{1}{2} \ln\frac{1 - t}{2} - \frac{3}{4}
\right)
- \frac{27}{4} - \frac{1 - t}{1 + t} \ln \frac{1 - t}{2}
\Bigg],
\nonumber \\
{^{G}\!T^{A(1)}}
\!\!\!&=&\!\!\!
\frac{N_f}{2} \left[
\left( \frac{1}{1 - t^2} + \frac{\ln\frac{1 - t}{2}}{(1 + t)^2} \right)
\left(
\ln\frac{1 - t}{2} - 2 \right)
- \frac{\ln^2 \frac{1 - t}{2}}{2(1 + t)^2} \right] .
\nonumber
\end{eqnarray}

%%%%%%%%%%%%%%%%%%%%%%%%%%%%%%%%%%%%%%%%%%%%%%%%%%%%%%%%%%%%%%%%%%%%%%%%%%%%
\section{Numerical estimations.}
%%%%%%%%%%%%%%%%%%%%%%%%%%%%%%%%%%%%%%%%%%%%%%%%%%%%%%%%%%%%%%%%%%%%%%%%%%%%

There are different proposals for modeling the SPD. Here we will deal only
with two of them, equating the SPD equal to the forward parton distributions
(FPD-model) and using the proposal of Radyushkin for the ansatz of the
double distribution (DD) functions \cite{MusRad99} $f ( z_-, z_+ ) =
\pi ( z_-, z_+ ) f ( z_+ )$, where the profile function $\pi$ for quarks
and gluons are given by
\begin{equation}
\label{profile}
{^{Q}\!\pi} (z_-, z_+)
= \frac{3}{4}
\frac{ [1 - |z_+|]^2 - z_-^2 }{ [1 - |z_+|]^3 },
\qquad
{^{G}\!\pi} (z_-, z_+)
= \frac{15}{16}
\frac{ \left\{ [1 - |z_+|]^2 - z_-^2 \right\}^2}{[1 - |z_+|]^5},
\end{equation}
The SPD is obtained by an integration \cite{MueRobGeyDitHor94}:
\begin{equation}
\label{DDtoSPD}
q (t, \eta, Q^2) = \int_{-1}^1 dz_+ \int_{- 1 + |z_+|}^{1 - |z_+|} dz_-
\delta ( z_+ + \eta z_- - t ) f ( z_-, z_+, Q^2 ) .
\end{equation}
We apply the DD model at different scales. To evolve the SPD, we expand
them:
\begin{eqnarray}
\label{bla}
q(t, \eta, Q^2) = \sum_{j = 0}^\infty c_j (\eta, Q^2) P_j (t)
\end{eqnarray}
in terms of Legendre polynomials and express the expansion coefficients
$c_j(\eta, Q^2)$ in terms of conformal moments which evolution is governed
by the anomalous dimension matrix (\ref{adcodt}). This method is suited
for not too small $\eta$.

%%%%%%%%%%%%%%%%%%%%%%%%%%%%%%%%%%%%%%%%%%%%%%%%%%%%%%%%%%%%%%%%%%%%%%%%%%%%
%                               Figure 2
%%%%%%%%%%%%%%%%%%%%%%%%%%%%%%%%%%%%%%%%%%%%%%%%%%%%%%%%%%%%%%%%%%%%%%%%%%%%
\begin{figure}[htb]
%\unitlength1mm
\centering
\inclfig{48}{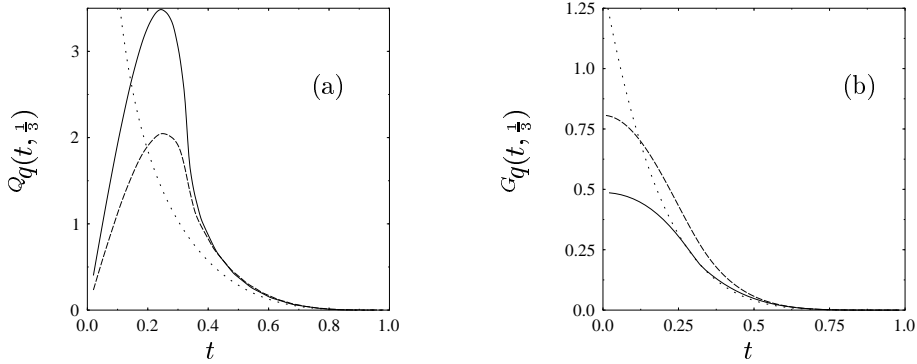}
%\begin{picture}(120,38)
%\put(-8,-18){\inclfig{12}{pic1a}}
%\put(40,25){(a)}
%\put(0,12){\rotate{${^Q\!q} (t, {\scriptstyle \frac{1}{3}})$}}
%\put(26,-10){$t$}
%\put(60,-18){\inclfig{12}{pic1b}}
%\put(110,25){(b)}
%\put(65,12){\rotate{${^G\!q} (t, {\scriptstyle \frac{1}{3}})$}}
%\put(97,-10){$t$}
%\end{picture}
\caption{
In (a) and (b) we show the SPD for the sum of the non-polarized $u$
and $\bar u$ quarks and the gluon densities, respectively. The
FPD-model (dotted line) and the DD-model with MRS
{\protect\cite{MarRobSti93}}
parametrization (dashed line) are modeled at a scale of $4\ {\rm GeV}^2$,
while the DD-Model with GRV {\protect\cite{GluReyVog98}} parametrization (solid
line) has been taken at the momentum scale $0.4\ {\rm GeV}^2$ and has
been evolved afterwards with NLO formulae to ${\cal Q}^2_0 = 4\ {\rm GeV}^2$.
The skewedness parameter is $\eta = 1/3$ which corresponds to $x_{\rm Bj} =
1/2$. \hfill\hfill
}
\label{FigInp}
\end{figure}
%%%%%%%%%%%%%%%%%%%%%%%%%%%%%%%%%%%%%%%%%%%%%%%%%%%%%%%%%%%%%%%%%%%%%%%%%%%%
%%%%%%%%%%%%%%%%%%%%%%%%%%%%%%%%%%%%%%%%%%%%%%%%%%%%%%%%%%%%%%%%%%%%%%%%%%%%
%                               Figure 3
%%%%%%%%%%%%%%%%%%%%%%%%%%%%%%%%%%%%%%%%%%%%%%%%%%%%%%%%%%%%%%%%%%%%%%%%%%%%
\begin{figure}[htb]
%\unitlength1mm
\centering
\inclfig{45}{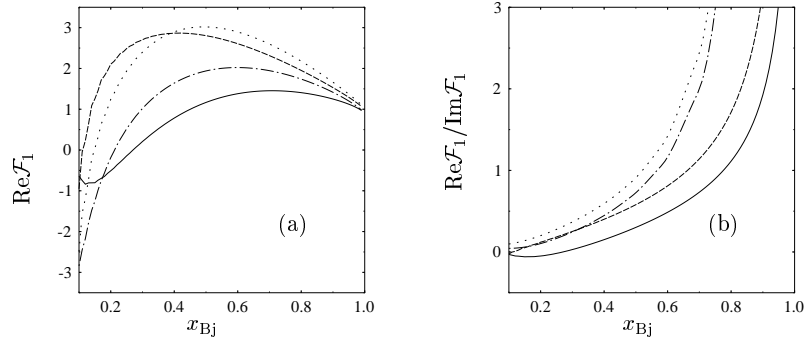}
%\begin{picture}(120,40)(0,0)
%\put(-10,-19){\inclfig{13}{pic2c}}
%\put(40,6){(a)}
%\put(26,-9){$x_{\rm Bj}$}
%\put(0,10){\rotate{${\rm Re} {\cal F}_1$}}
%\put(55,-19){\inclfig{13}{pic2b}}
%\put(105,6){(b)}
%\put(92,-9){$x_{\rm Bj}$}
%\put(65,12){\rotate{${\rm Re} {\cal F}_1/{\rm Im} {\cal F}_1$}}
%\end{picture}
\caption{
The real part (a) of ${\cal F}_1$ for DD-model with MRS and GRV
parametrizations for the scale ${\cal Q}_0^2=4\ {\rm GeV}^2$ and for $x_{\rm
Bj} \ge 0.1$. Here MRS (GRV) amplitude is plotted at LO as dashed (dotted)
and at NLO as solid (dash-dotted) lines, respectively.
The ratio of the real to imaginary part (a) of
${\cal F}_1$, respectively, for the DD- and FPD-model with the MRS
parametrization. Here the DD(FPD)-model at LO and NLO is shown as
dashed (dotted) and solid (dash-dotted) lines, respectively.
For the SPD-models considered the result is independent of $\Delta^2$.
\hfill\hfill
}
\label{pred1}
\end{figure}
%%%%%%%%%%%%%%%%%%%%%%%%%%%%%%%%%%%%%%%%%%%%%%%%%%%%%%%%%%%%%%%%%%%%%%%%%%%%

In Fig.\ \ref{FigInp} we show our input distributions at a scale of ${\cal
Q}^2 = -q_1^2 =4 \mbox{GeV}^2$. Let us also mention, that the relative NLO
corrections due to the evolution are as expected not very large. Starting at
an input scale of 0.7 GeV, we found that the NLO corrections do not exceed
10\% to 30\% \cite{BelMueNieSch98BelMueNieSch98a}.

For the numerical estimation of the hard-scattering amplitudes (\ref{def-T})
we perform also an expansion with respect to Legendre polynomials which
allows us to include easily evolution effects. However, this method is not
appropriate to study the low $x_{\rm Bj}$ region. In general we found that
the NLO corrections for $x_{\rm Bj} > 0.1$ are typically of the order of
20-50\% (see Fig.\ \ref{pred1}a). In special cases, e.g. for the DD model
with MRS parameterization, it can even be much larger. Fortunately, there is
partial cancellation of perturbative corrections in the ratio of real to
imaginary part. As shown in Fig.\ \ref{pred1}b, this ratio is model
dependent and may its measurement can be used to get a deeper insight in the
SPD.

A.B. was supported by the Alexander von Humboldt foundation.

\end{document}